\documentclass[submission, PhysProc]{SciPost}

\hypersetup{
    colorlinks,
    linkcolor={red!50!black},
    citecolor={blue!50!black},
    urlcolor={blue!80!black}
}

\usepackage[bitstream-charter]{mathdesign}
\DeclareSymbolFont{usualmathcal}{OMS}{cmsy}{m}{n}
\DeclareSymbolFontAlphabet{\mathcal}{usualmathcal}

\begin{document}

% TODO: write your article's title here.
% The article title is centered, Large boldface, and should fit in two lines
\begin{center}{\Large \textbf{
Three-Nucleon Force Effects in the FSI configuration of the
   $d (n,nn) p$ Breakup Reaction
\\
}}\end{center}

% TODO: write the author list here. Use first name (+ other initials) + surname format.
% Separate subsequent authors by a comma, omit comma and use "and" for the last author.
% Mark the corresponding author with a superscript star.
\begin{center}
Hiroyuki Kamada \textsuperscript{1,$\star$},
Henryk Wita\l a \textsuperscript{2},
Jacek Golak     \textsuperscript{2} and
Roman Skibi\'nski \textsuperscript{2} 
%Gee K. See\textsuperscript{3$\star$}
\end{center}

% TODO: write all affiliations here.
% Format: institute, city, country
\begin{center}
{\bf 1} 
Department of Physics, Faculty of Engineering, Kyushu Institute of Technology, \\
Kitakyushu 804-8550, Japan
\\
{\bf 2} %Affiliation2
M. Smoluchowski Institute of Physics, Jagiellonian University, 
%PL-
30348 
Krak\'ow, Poland %, \email{henryk.witala@uj.edu.pl
\\
%{\bf 3} Affiliation2
%\\
% TODO: provide email address of corresponding author
${}^\star$ {\small \sf kamada@mns.kyutech.ac.jp}
\end{center}

\begin{center}
\today
\end{center}

% For convenience during refereeing (optional),
% you can turn on line numbers by uncommenting the next line:
%\linenumbers
% You should run LaTeX twice in order for the line numbers to appear.

\definecolor{palegray}{gray}{0.95}
\begin{center}
\colorbox{palegray}{
  \begin{tabular}{rr}
  \begin{minipage}{0.05\textwidth}
    \includegraphics[width=14mm]{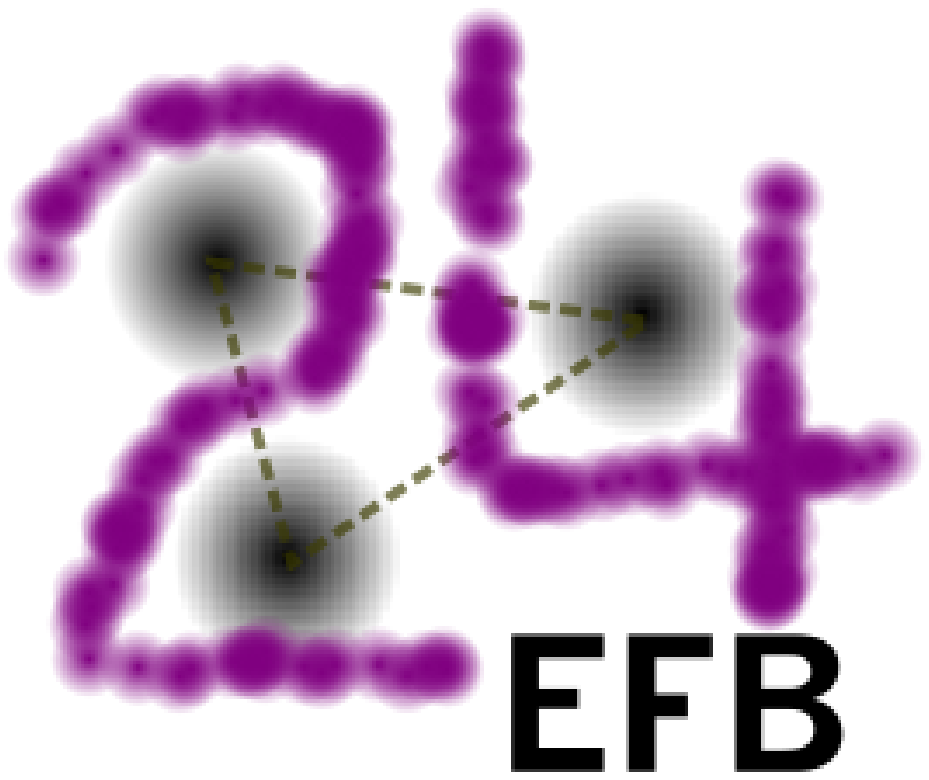}
  \end{minipage}
  &
  \begin{minipage}{0.82\textwidth}
    \begin{center}
    {\it Proceedings for the 24th edition of European Few Body Conference,}\\
    {\it Surrey, UK, 2-4 September 2019} \\
    %\doi{10.21468/SciPostPhysProc.2}\\
    \end{center}
  \end{minipage}
\end{tabular}
}
\end{center}

\section*{Abstract}
{\bf
% TODO: write your abstract here.
We investigated three-nucleon (3N) force effects in the final state
  interaction (FSI) configuration of the $d (n,nn) p$ breakup reaction
at the incoming nucleon energy $E_n$= 200 MeV. 
  Although 3N force effects for the elastic nucleon-deuteron
scattering cross section at comparable energies
are located predominantly in the region of intermediate and backward angles,
 the corresponding 3N force effects for the integrated 
FSI configuration breakup 
cross section are found also at forward scattering angles.
}

% TODO: include a table of contents (optional)
% Guideline: if your paper is longer that 6 pages, include a TOC
% To remove the TOC, simply cut the following block
%\vspace{10pt}
%\noindent\rule{\textwidth}{1pt}
%\tableofcontents\thispagestyle{fancy}
%\noindent\rule{\textwidth}{1pt}
%\vspace{10pt}

\section{Introduction}
\label{sec:intro}
% TODO: write your article here.
Our studies of 3N continuum are based on the exact solutions 
of the 3N Faddeev equation in momentum space. They began in the 1980s 
and in the 1990s were performed with several realistic two-nucleon (2N) forces:
the AV18~\cite{Wiringa:1994wb}, CD Bonn~\cite{Machleidt:2000ge}, 
NijmI, NijmII, Nijm93 and Reid93~\cite{Stoks:1994wp} potentials.
The results of these studies (see for example 
Refs.~\cite{Gloeckle:1995jg,Witala:2000am}) proved 
that predictions of 3N scattering observables are in good agreement 
with the data at input nucleon energies below about 30 MeV.
The situation changed at highier energies, 
where theoretical predictions using only 2N forces 
clearly deviated from the data~\cite{Sakai:2000mm,Kistryn:2003xs}.
In particular, strong discrepancies between such calculations based on 2N 
potentials and the data were found in the minimum of the elastic scattering 
cross section. For energies smaller than approximately 140 MeV
the agreement between theoretical predictions and the data for 
this observable was regained, when the Tucson–Melbourne (TM)~\cite{Coon:2001pv} 
or Urbana IX~\cite{Pudliner:1997ck} 3N force (3NF) models were included 
in the 3N Hamiltonian~\cite{Witala:1998ey}. Thus the studies 
in Ref.~\cite{Witala:1998ey} provided strong evidence for the action of 3NF 
in 3N scattering. However, the description of many polarization observables
and generally the description of the data at still higher energies 
was not always satisfactory~\cite{Witala:2000am,Bieber:2000zz}.

At high energies one could expect deficiencies in the nonrelativistic 
Faddeev approach. That is why we constructed a relativistic framework 
in the form of relativistic Faddeev equations
~\cite{Kamada:1999wy,Kamada:1999fz,Kamada:2007ms,Witala:2011yq,Polyzou:2010kx,Sekiguchi:2005vq,Witala:2008va} 
%~\cite{Kamada:1999wy,Kamada:1999fz,Witala:2011yq,Polyzou:2010kx,Sekiguchi:2005vq,Witala:2008va} 
according to the  Bakamjian-Thomas theory~\cite{Bakamjian:1953kh}. However, the relativistic 
effects turned out to be generally small and insufficient to significantly
improve the data description. 

Neither TM nor Urbana IX could be considered merely as phenomenological
3NF models, since they are based on a meson theoretical picture. 
However, it was pointed out that these 3N forces were not consistent 
with the widely used 2N forces. The QCD Lagrangian with massless quarks 
possesses chiral symmetry. This chiral symmetry is explicitly 
broken because of the quark mass terms. 
This feature of QCD and the mechanism of spontaneous chiral symmetry breaking 
inspired Weinberg to use effective field theory of QCD in the form 
of chiral perturbation theory as a tool to construct nuclear interactions.
This idea was then implemented by many physicists, who strove for construction 
of precision 2N and many-nucleon potentials. 
We mention here work by van Kolck~\cite{vanKolck:1994yi}, 
the early model of the Bochum-Bonn group~\cite{Epelbaum:1998ka} 
and the nuclear forces developed by the Moscow (Idaho)-Salamanca 
group~\cite{Entem:2001qp}. In particular Epelbaum and collaborators for the
first time used chiral 2N and 3N forces to study nucleon-deuteron 
scattering~\cite{Epelbaum:2002vt}.

Currently the investigations of few- and many-nucleon systems 
with the new generations of chiral potentials from the Bochum-Bonn group 
are carried out within the LENPIC project~\cite{LENPIC:2019lp}. 
More information about this initiative, coordinated by E. Epelbaum and 
J. Vary, can be found in the contribution to this conference 
by J. Golak et al.~\cite{GOLAK:2019efb24}.

In the present contribution we studied
in detail one of the most important kinematical configurations
of the nucleon-induced deuteron breakup reaction, namely the final 
state interaction (FSI) configuration. 
We considered the case, where two neutrons emerged with the same momenta,
forming quasi dineutron, while the final proton momentum was restricted 
by four-momentum conservation. Our purpose was to estimate 3NF effects 
for this effectively two-body reaction.

\section{Final State Interaction configuration}

We investigated 3NF effects in the FSI configuration of the 
$d (n,nn) p$ breakup reaction. To this end we obtained 
solutions of the 3N Faddeev equations~\cite{Gloeckle:1995jg} with the CDBonn
nucleon-nucleon potential \cite{Machleidt:2000ge} and the 
Tucson-Melbourne 3NF \cite{Coon:2001pv}. From these solutions one 
can construct not only the elastic scattering observables but also 
the observables for the breakup process.
In this contribution we restrict ourselves to an integrated breakup 
cross section around the final state interaction condition for 
the two emerging neutrons: 
\begin{eqnarray}
{d ^2 \sigma \over d\Omega_1 d\Omega_2 } \equiv \left. \int _{S_0-\Delta S}^{S_0+\Delta S} 
{d^3 \sigma \over d\Omega_1 d\Omega_2 dS}  dS \right|_{\Omega_1 = \Omega_2}
~~~~~ ( n + d \to (nn) + p )
\label{eq.1}
\end{eqnarray}
Here $\Omega_1$ and $\Omega_2$ represent the directions of the momenta 
of the outgoing neutrons 1 and 2, respectively. 
Note that for fixed $\Omega_1$ and $\Omega_2$, the energies of the two 
neutrons, $E_1$ and $E_2$ lie on a certain curve, the so-called 
"kinematical locus". 
Choosing an appropriate starting point 
where by definition $S=0$, the $S$ parameter is calculated 
as a distance taken along the curve from its starting point:  
\begin{eqnarray}
S= \int dS = \int \sqrt {(dE_1)^2 + (dE_2)^2 } \, .
\end{eqnarray}
This arc-length variable $S$ defines uniquely the three-nucleon kinematics, 
yielding a specific $(E_1,E_2)$ point on the kinematically allowed
curve in the $(E_1,E_2)$ plane.
The FSI occurs for the condition $E_1=E_2$, where $S \equiv S_0$.

\begin{figure}[hbpt]
\vspace{-0.5cm}
\begin{center}
\includegraphics[scale=0.4]{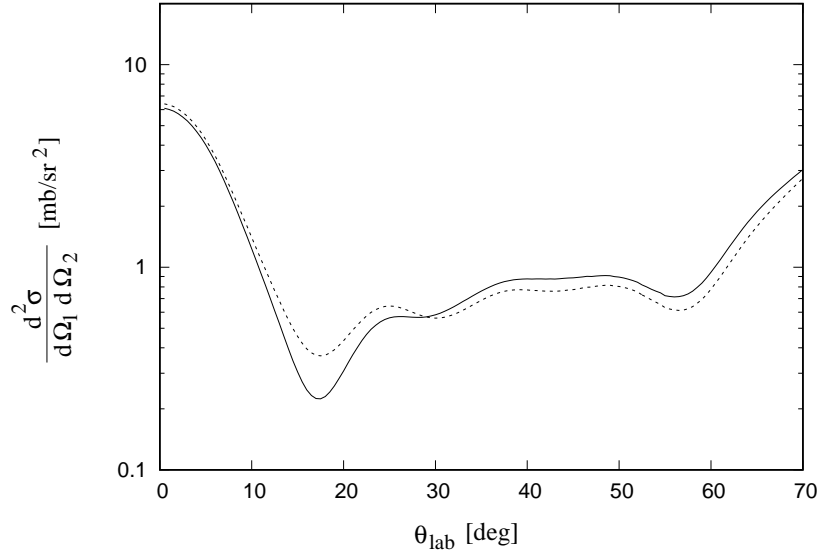}
\caption{The integrated final state interaction configuration
breakup cross section for 
the incident neutron laboratory kinetic energy $E_n$= 200~MeV.  
The theoretical predictions based solely on a 2N interaction 
(here the the CDBonn potential \cite{Machleidt:2000ge}) are represented by the dotted 
line, while the results obtained with the 2N potential augmented by 
the Tucson-Melbourne 3NF \cite{Coon:2001pv} are shown with the solid line.
\label{fig1}
}
\end{center}
\end{figure}

Figure \ref{fig1} shows the breakup cross section 
of Eq.~(\ref{eq.1}) for the incident neutron laboratory energy $E_{n}$=200~MeV
resulting from integrations over the $S$ variable
in the interval $S_0 - \Delta S , S_0 + \Delta S $
with the width parameter $\Delta S$= 20~MeV. 
The angle $\theta_{lab}$ is the common laboratory scattering angle
of nucleons 1 and 2, for which the FSI condition is realized.

We found a large deviation between the theoretical predictions for the FSI cross section including or not including 3NF.
Although the 3NF effects for the elastic scattering cross section 
are located  predominantly  in the region starting from  middle up to backward scattering angles,
 the 
 3NF effects for the integrated FSI configuration breakup cross section are
  found also at forward scattering angles.
%\keywords{Faddeev 3-body calculation, nucleon-deuteron breakup scattering}

\section{Conclusion}

We found a large deviation between the theoretical predictions including or not including 3NF.
Although the 3NF effects for the elastic scattering cross section 
\cite{Witala:1998yd,Sakai:2000mm,Kistryn:2003xs,Maeda:2007zza}
are most pronounced for the intermediate and backward scattering angles,
the 3NF effects for the integrated FSI configuration breakup cross section 
are found also at forward scattering angles. 
We hope that our results can be
in future confronted with experimental data.
%However, in this case, certainly it is not easy to set up such an experiment 
%to detect two neutrons using a neutron beam. 
%Considering that in the final state, the two neutrons have 
%the same scattering angle, it needs actually be measured 
%by capturing the remaining protons.

\section*{Acknowledgements}
The numerical calculations were partially
performed on the interactive server at RCNP, Osaka University, Japan, 
and on the supercomputer cluster of the JSC, J\"ulich,
Germany.

% TODO: include funding information
\paragraph{Funding information}
%Authors are required to provide funding information, including relevant agencies and grant numbers with linked author's initials. Correctly-provided data will be linked to funders listed in the \href{https://www.crossref.org/services/funder-registry/}{\sf Fundref registry}.
This work was supported by the Polish National Science Centre under Grants No. 2016/22/M/ST2/00173 
and 2016/21/D/ST2/01120.% 

\nolinenumbers

\end{document}